\documentclass{article}
\usepackage{arxiv}
\usepackage[utf8]{inputenc} 
\usepackage[T1]{fontenc}    
\usepackage{hyperref}
\usepackage{ upgreek }
\usepackage{url}            
\usepackage{booktabs}       
\usepackage{amsfonts}       
\usepackage{nicefrac}       
\usepackage{microtype}      
\usepackage{lipsum}
\usepackage{float}
\usepackage{graphicx}
\hypersetup{
    colorlinks=true,
    linkcolor=blue,
    filecolor=magenta,      
    urlcolor=cyan,
    }
\usepackage{multirow}
\usepackage{array}
\usepackage{comment}
\usepackage{longtable}
\usepackage{ragged2e}
\usepackage{subfig}
\usepackage{amsmath}
\usepackage{longtable}
\usepackage{ amssymb }
\title{Introduction of Medical Imaging Modalities}

\author{
  S. K. M Shadekul Islam\\
  Research and Development Department, Pioneer Alpha,\\
Dhaka, Bangladesh\\
  \texttt{shadekul15-12836@diu.edu.bd} \\
   \And
    MD Abdullah Al Nasim   \\
  Research and Development Department, Pioneer Alpha,\\
Dhaka, Bangladesh\\
  \texttt{nasim.abdullah@ieee.org}\\
   \And
      Ismail Hossain \\
   University of Alabama at Birmingham\\
Alabama, USA\\
  \texttt{ihossain@uab.edu} \\
   \And
   Dr. Md Azim Ullah \\
   University of Memphis\\
  \texttt{mullah@memphis.edu} \\
  \And
Dr. Kishor Datta Gupta\\
Clark Atlanta University\\
Georgia, USA\\
\texttt{kgupta@cau.edu}
 \And
    Md Monjur Hossain Bhuiyan\\
  School of Aerospace and Mechanical Engineering\\
University of Oklahoma\\
Norman, OK-73019, USA\\
\texttt{mhbhuiyan@ou.edu}\\
}

\begin{document}
\maketitle

\begin{abstract}
The diagnosis and treatment of various diseases had been expedited with the help of medical imaging. Different medical imaging modalities, including X-ray, Computed Tomography (CT), Magnetic Resonance Imaging (MRI), Nuclear Imaging, Ultrasound, Electrical Impedance Tomography (EIT), and Emerging Technologies for in vivo imaging are widely used in today’s healthcare throughout the world. A review of medical imaging modalities is presented in this chapter, in addition to these modalities, some advanced techniques such as contrast-enhanced MRI, MR approaches for osteoarthritis, Cardiovascular Imaging, and Medical Imaging data mining and search. Despite its important role and potential effectiveness as a diagnostic tool, reading and interpreting medical images  by radiologists is often tedious and difficult due to the large heterogeneity of diseases and the limitation of image quality or resolution. Besides the introduction and discussion of the basic principles, typical clinical applications, advantages, and limitations of each modality used in current clinical practice, this chapter also highlights the importance of emerging technologies in medical imaging and the role of data mining and search aiming to support translational clinical research, improve patient care, and increase the efficiency of the healthcare system.

\end{abstract}

\section{Introduction}
Medical imaging is a very essential tool in the diagnosis and treatment of various diseases. It refers to the non-invasive techniques and approaches used to create visual representations of the internal organs and tissues of the human body. These visual representations, also known as images, can be used to detect and diagnose a variety of diseases, guide disease treatment strategies and monitor treatment efficacy. Specifically, medical imaging is widely used to examine and visualize different parts of
the body, including bones, muscles, organs, blood vessels, and other internal structures. Medical imaging can also be divided into two main categories: diagnostic imaging and therapeutic imaging \cite{Imaginga41:online}. Diagnostic imaging, which is used to detect and diagnose diseases, and determine disease severity, include modalities such as X-ray radiography, computed tomography (CT), magnetic resonance imaging (MRI), ultrasound (US), and nuclear medicine. Therapeutic imaging, which is used to guide procedures such as surgery or radiation therapy, includes modalities such as fluoroscopy, angiography, and interventional radiology. With the advancements in technology, a wide range of medical imaging modalities have been developed to provide detailed visual information about the internal structures and functions of the body. 

In current clinical practice, the routinely used imaging modalities include X-ray, CT, MRI, ultrasound, and nuclear medicine. Each imaging modality has its unique advantages for different applications and diagnostic purposes, and limitations to reveal the internal structure of the different organs of the body and their functions \cite{hossainbrain}. This chapter aims to provide an overview of the most commonly used medical imaging modalities, including X-ray, CT, MRI, ultrasound, and nuclear medicine. We will discuss the basic principles, typical clinical applications, advantages, and limitations of each modality. Additionally, we will explore the latest developments and future directions in the field of medical imaging. The goal of this chapter is to provide a comprehensive understanding of the various medical imaging modalities available today and their impact on the field of medicine.

\section{Background Study}
Medical imaging modalities refer to the various techniques and technologies used to create visual representations of the human body for disease diagnosis and treatment purposes. In recent years we have seen a good deal of research to further improve the efficacy and image quality of many existing medical imaging modalities including X-ray, CT, MRI, ultrasound, nuclear medicine, cardiovascular imaging, electrical impedance tomography (EIT), and positron emission tomography (PET). 
\\\\
Additionally, Chapman's (2009) X-ray imaging, also known as radiography, is one of the oldest and most widely used imaging modalities. It is based on the principle that different types of tissue absorb different amounts of X-ray radiation, allowing for the creation of detailed images of the body's internal structures~\cite{Chapman2009}. However, X-ray imaging has some limitations, such as exposure to ionizing radiation and the lack of tissue contrast.
\\\\
Ole Petter Wennberg et. al 2018 \cite{doi:10.1144/SP459.10} CT scans, also known as computed tomography, use X-ray radiation to create detailed cross-sectional images of the body. CT scans are highly effective in detecting bone and soft tissue abnormalities, and they can also be used to guide biopsies and other procedures. However, CT scans also involve exposure to ionizing radiation and may not be suitable for certain patient populations.
\\\\
Paul Glover 2013 \cite{Glover2013} MRI uses a magnetic field and radio waves to create detailed images of the body's internal structures. MRI is particularly effective at imaging soft tissue and is often used to detect tumors, injuries, and other abnormalities in the brain, spine, and other parts of the body. MRI is non-invasive and does not involve ionizing radiation, making it a safer option for certain patients.
\\\\
Yobo Xie 2021 \cite{Xie:2021:2156-7018:930} Ultrasound imaging uses high-frequency sound waves to create images of the body's internal structures. It is particularly effective at imaging soft tissue and is often used to image the fetus during pregnancy, as well as to evaluate organs such as the liver, gallbladder, and kidneys. Ultrasound is non-invasive, does not involve ionizing radiation, and has no known harmful effects.
\\\\
Sandip Basu et. al 2011 \cite{https://doi.org/10.1111/j.1749-6632.2011.06077.x} PET imaging uses small amounts of radioactive material to create detailed images of the body's metabolic processes. It is often used in conjunction with other imaging modalities, such as CT and MRI, to create more detailed images of certain diseases and conditions. However, PET imaging involves exposure to ionizing radiation and may not be suitable for certain patient populations.
\\\\
In conclusion, various medical imaging modalities are available, each with its own advantages and disadvantages. X-ray, CT, MRI, ultrasound, and PET are some of the most commonly used modalities. X-ray and CT use ionizing radiation to create images, which can be harmful to the patient, while MRI and ultrasound use non-ionizing radiation, which is safer. PET uses ionizing radiation, but it is used in conjunction with other imaging modalities to create more detailed images of certain diseases and conditions \cite{tonmoy2019brain}.

\section{Methodology}
First of all in the research methodology for this chapter on "Introduction of Medical Imaging Modalities" I identified the different medical imaging modalities that will be covered in the paper such as X-ray, CT Scan, MRI, Ultrasound imaging, Nuclear medicine imaging, Electrical Impedance Tomography, Cardiovascular Imaging. The paper is focused on introducing these modalities in terms of their basic working principle, applications, limitations, and advantages.
\\\\
In the second step, we conducted a comprehensive literature search using relevant databases such as Google, Google Scholar, PubMed, Scopus, and Web of Science using keywords such as "Medical Imaging Modalities", "X-ray", "CT Scan", "MRI", "Ultrasound imaging", "Nuclear medicine imaging", "Electrical Impedance Tomography", "Cardiovascular Imaging" used to find relevant studies. Inclusion/exclusion criteria such as the date range of the studies and the language of the studies are applied to narrow down the search.
\\\\
Then thirdly, we carefully review and assess the quality of the studies found in the literature search, and select the most relevant and high-quality studies for the review. The selected studies covered the basic working principle, applications, limitations, and advantages of the different medical imaging modalities. Relevant information is extracted from these studies. The results are presented in a clear and concise manner, including a discussion of the strengths and limitations of the studies.
\\\\
In the fourth step, we provided a conclusion that summarizes the main findings of the review. This section provides a clear summary of the key takeaways from the review.
\\\\
And finally, we included a reference list of all the studies that were included in the review, following the Springer citation style. This is an important step that allows other researchers to easily find and access the studies that were used in the review.
\\\\
Overall, the methodology for this review paper on "Introduction of Medical Imaging Modalities" is a systematic and thorough process that will help to ensure that the review is comprehensive, accurate, and up-to-date. \\

\begin{figure}
    \centering
    \includegraphics[width=1\textwidth]{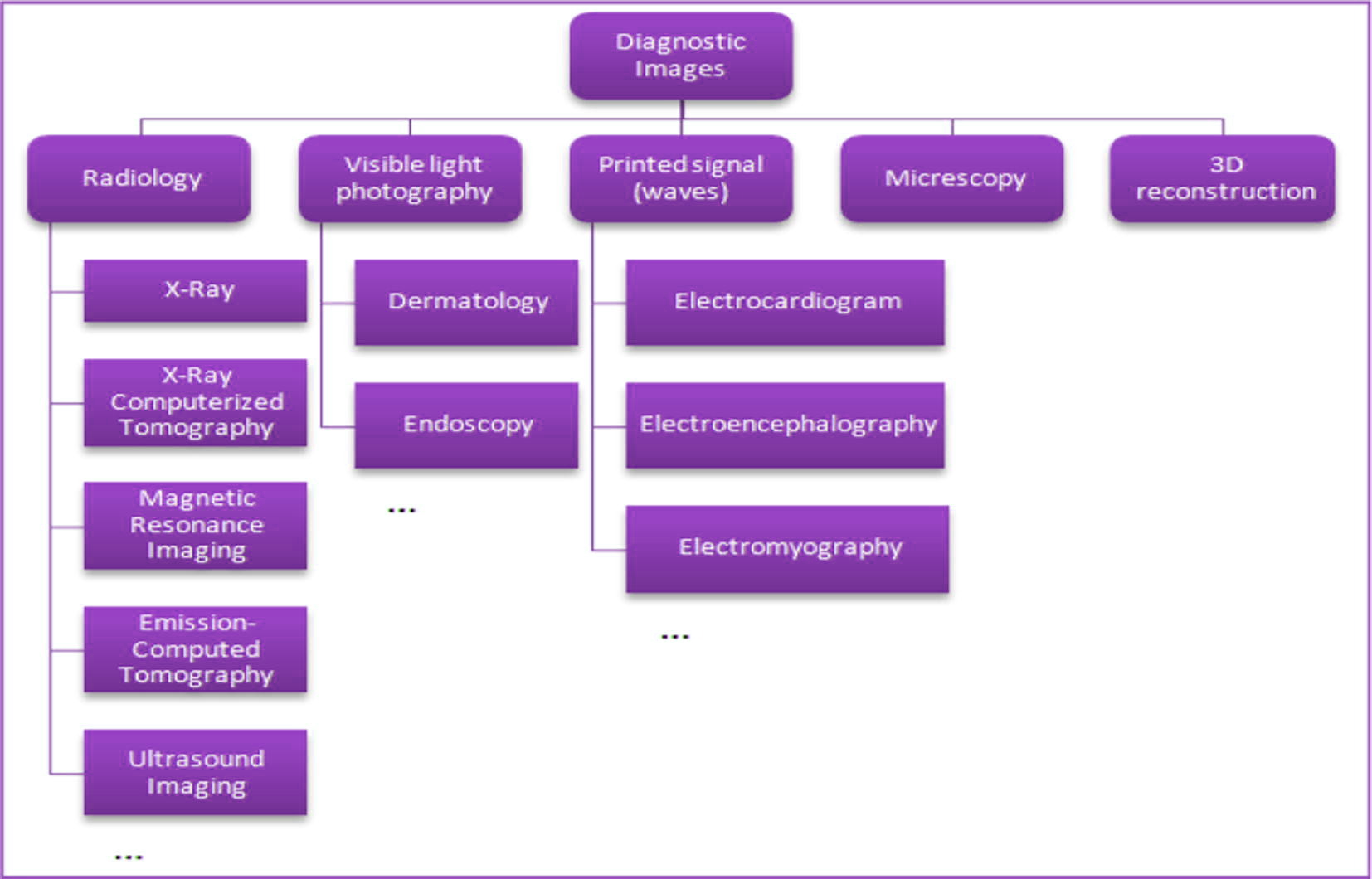}
    \caption{The categorization of medical imaging modalities \cite{ELGAMAL201699}.}
    \label{fig1}
\end{figure}

Figure \ref{fig1} depicts the five groups into which the diagnostic pictures are divided. As a result of the various medical technology each of these groups employs, the output images they create vary. Although all of the images fall under the same general category structure, each one has unique qualities that aid in the production of various kinds of information. The information generated by various picture modalities to deal with registration and/or fusion procedures is described in this section. In accordance with the kind of output photos, this part is divided into five subsections. These divisions are X-ray imaging, visible light photography, printed signals (waves), microscopy, and three-dimensional representation.

\section{Definition of medical imaging}
Medical imaging refers to the techniques and technologies used to create visual representations of the inside of the body. \cite{ahmad2014medical} These visual representations, also known as images, can be used to diagnose, monitor, or treat various medical conditions. Medical imaging can be used to examine and visualize different parts of the body, including bones, muscles, organs, blood vessels, and other internal structures.
\\\\
Medical imaging can be divided into two main categories: diagnostic imaging and therapeutic imaging \cite{Imaginga41:online}. Diagnostic imaging, which is used to diagnose and determine the severity of a condition, includes modalities such as X-ray, CT, MRI, ultrasound, and nuclear medicine. Therapeutic imaging, which is used to guide procedures such as surgery or radiation therapy, includes modalities such as fluoroscopy, angiography, and interventional radiology.
\\\\
Some common modalities include X-ray, CT, MRI, ultrasound, and nuclear medicine, each has its own indications and limitations, these modalities are powerful diagnostic tools that can reveal the internal structure of the body and its functions \cite{ali2021enhanced}.

\section{Overview of different modalities}

\subsection{X-ray imaging:} X-rays are a form of electromagnetic radiation that can pass through many solid objects, including the human body. X-ray imaging uses X-rays to create detailed images of internal structures, such as bones \cite{XRayImag53:online}.
\subsection{Computed Tomography (CT) imaging:} CT scans use X-rays to create detailed cross-sectional images of the body. These images can be used to diagnose and monitor a wide range of conditions, including cancer, heart disease, and injuries to the bones and internal organs.\cite{Buzug2011}
\subsection{Magnetic Resonance Imaging (MRI) imaging:} With the use of a strong magnetic field, radio waves, and a computer, MRI scans can produce very detailed images of internal structures, such as the brain and spinal cord, muscles, tendons, ligaments, and blood vessels.\cite{Glover2013}
\subsection{Ultrasound imaging:} Ultrasound employs acoustic waves of very high frequency to create images of internal structures, such as the uterus and ovaries during pregnancy. It is also used to examine the organs and blood vessels in the abdomen and to guide needle biopsies.\cite{ahmad2014medical}
\subsection{Nuclear medicine imaging:} Nuclear medicine uses small amounts of radioactive material to create images of internal structures and to identify abnormalities, such as tumors or infections. This modality is also used to diagnose and monitor certain diseases such as cancer, heart diseases, and thyroid diseases.\cite{NuclearM94:online}
\subsection{Electrical Impedance Tomography (EIT):} EIT imaging uses electrical signals to image internal structures and monitors changes in their electrical properties, such as the lungs and heart.\cite{doi:10.1080/0309190021000059687}
\subsection{Cardiovascular Imaging:} It is a subspecialty of Medical imaging that uses various modalities to visualize the structure and function of the heart and blood vessels. It includes techniques such as Echocardiography, Cardiac Computed Tomography (CCT), and Magnetic Resonance Imaging (CMR). These techniques help in the diagnosis of heart diseases, such as Coronary artery disease, valvular heart disease, and congenital heart disease. They also play an important role in guiding interventional procedures such as angioplasty and stent placement.

\vspace{5mm}
\section{X-ray Imaging}
\subsection{Basic principles}
X-ray imaging, often known as radiography, is an imaging technique used in medicine for diagnosis that creates high-resolution images of internal anatomical structures like bones. The basic principles of X-ray imaging are as follows:
\\\\
X-rays are a form of electromagnetic radiation that can pass through many solid objects, including the human body. This type of imaging works by directing a beam of X-rays at the area of the body being examined.\cite{Chapman2009} The X-rays pass through the body and are then detected by a special detector, such as an X-ray film or digital detector. Because bones are denser than other tissues, they absorb more X-rays and appear white on the final image, while softer tissues such as muscles and organs absorb fewer X-rays and appear darker. In order to produce a clear image, the machine must use the appropriate amount of radiation and the body part must be positioned correctly in relation to the X-ray beam. The image produced is a two-dimensional representation of the three-dimensional structure of the body, so to obtain a more complete picture, different views are taken to examine the same area from different angles. The images are then interpreted by a radiologist, who is a medical doctor who specializes in interpreting medical images, to identify any abnormalities or issues, such as broken bones, tumors, or other conditions, and provide a diagnosis.
\begin{figure}
    \centering

    \includegraphics[width=.6\textwidth]{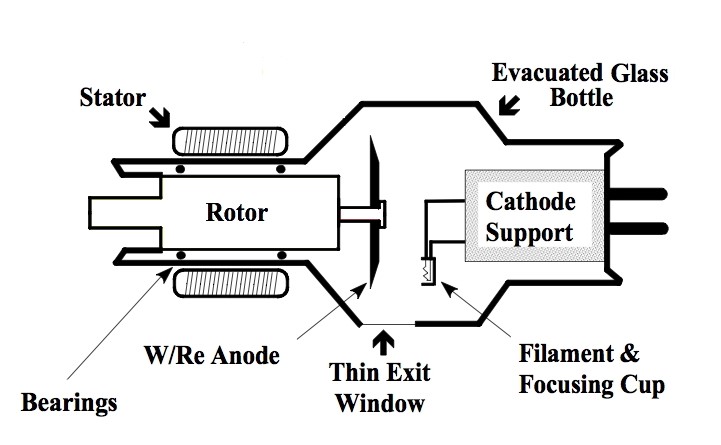}
    \caption{X-ray Imaging Principle\cite{Xrays–Un32:online}}
\end{figure}

Electromagnetic radiation, or X-rays, are created in an evacuated glass tube (the bottle in Figure 2) \cite{Xrays–Un32:online}. High-speed electrons that have been propelled across the tube's vacuum by a significant voltage difference between the cathode and the anode are aimed at a rotating tungsten anode.  X-rays and heat are released when the electrons hit the anode.  The X-ray tube is inactive until the Medical Radiation Technologist (MRT) turns it on because X-rays can only be created when a voltage differential is introduced across the cathode and anode. All imaging modalities that utilise X-rays (X-ray, CT, Fluoroscopy, Angiography) use this physical, on/off configuration.

\subsection{Typical Clinical Applications}
X-ray imaging, also known as radiography, has a wide range of clinical applications in medicine.\cite{epstein2007introduction} Some typical clinical applications of X-ray imaging include:
\\\\
X-ray imaging is a widely used medical tool that has a variety of applications. It is commonly used to examine bones for signs of fractures, dislocations, or other injuries. It can also detect bone diseases such as osteoporosis, osteoarthritis, and bone tumors. In addition to bone imaging, X-ray imaging is also used to evaluate the chest and detect lung conditions such as pneumonia, lung cancer, and emphysema. It can also be used to evaluate the size, shape, and position of the heart and blood vessels. X-ray imaging can also be used to examine the abdominal organs such as the liver, spleen, and kidneys, as well as detect conditions such as gallstones, kidney stones, and intestinal obstruction. In dentistry, X-ray imaging is commonly used to detect tooth decay, abscesses, and impacted teeth, as well as evaluate the jaw and temporomandibular joint. In emergency departments and trauma centers, X-ray imaging is used to quickly and accurately diagnose injuries such as fractures, dislocations, and foreign bodies. Additionally, X-ray imaging can be used to monitor the progress of treatment for certain conditions, such as monitoring the healing of a fracture or the response of a tumor to radiation therapy.
\\\\
X-ray imaging, also known as radiography, is a widely used and valuable diagnostic tool in medicine. Some advantages of X-ray imaging include:
\\\\
X-ray imaging is a popular diagnostic tool because of its wide availability, speed, and low cost. X-ray imaging equipment is widely available in hospitals, clinics, and doctors' offices, making it a convenient and accessible diagnostic tool. Additionally, X-ray imaging can be performed quickly, and results are usually available within a short period of time. Moreover, X-ray imaging is relatively inexpensive compared to other imaging modalities such as MRI or CT. Another advantage is that X-ray imaging produces high-resolution images that can reveal fine details of bone and other internal structures. Additionally, X-ray imaging is a non-invasive technique, meaning that it does not involve the use of incisions or injections, making it relatively safe and comfortable for the patient.
\\\\
However, there are also some limitations of X-ray imaging to consider:
\\\\
X-ray imaging has some limitations, one of them being limited tissue contrast. It is not as good as other modalities at distinguishing between different types of tissue, such as tumors and normal tissue. Additionally, X-ray imaging has low soft tissue contrast, meaning it is not as good as other modalities at visualizing soft tissue, such as muscles and organs. Another limitation is that X-ray imaging produces a 2D representation of the 3D structure of the body, which can make it difficult to visualize certain internal structures, particularly in areas that are not easily accessible. This 2D representation can make it difficult to understand some areas of internal structures and make it hard to diagnose certain issues.
\vspace{5mm}
\section{Computed Tomography (CT) Imaging}
\subsection{Basic principles}
CT imaging, commonly called CAT scanning due to its use of X-rays to make cross-sectional pictures of the body, is a common medical imaging procedure. The basic principles of CT imaging are as follows:
\\\\
CT scans use X-rays to create detailed images of internal structures by directing the X-rays at the body from different angles and measuring the intensity of the X-rays that pass through the body with detectors. A special type of X-ray detector called a multi-slice detector is used in CT scans, which can acquire multiple images at the same time from different angles, allowing for the creation of detailed cross-sectional images of the body. To improve image quality, CT scans also use techniques such as spatial filtering, which removes noise and improves contrast, and multi-energy imaging, which uses different energy levels of X-ray beams to capture different information and increase image contrast. Additionally, CT scans can use a technique called dose modulation, which adjusts the radiation dose according to the size, shape, and composition of the body part being scanned, reducing the risk of side effects. The images obtained from CT scans are then processed by a computer and can be viewed in different ways, such as cross-sectional slices, 3D images, and even virtual reality images \cite{BROOKS1993575, islam2021brain, nasim2021prominence}. CT scans are performed by radiologic technologists and the images are interpreted by radiologists, who are medical doctors who specialize in interpreting medical images.
\begin{figure}
    \centering

    \includegraphics[width=.6\textwidth]{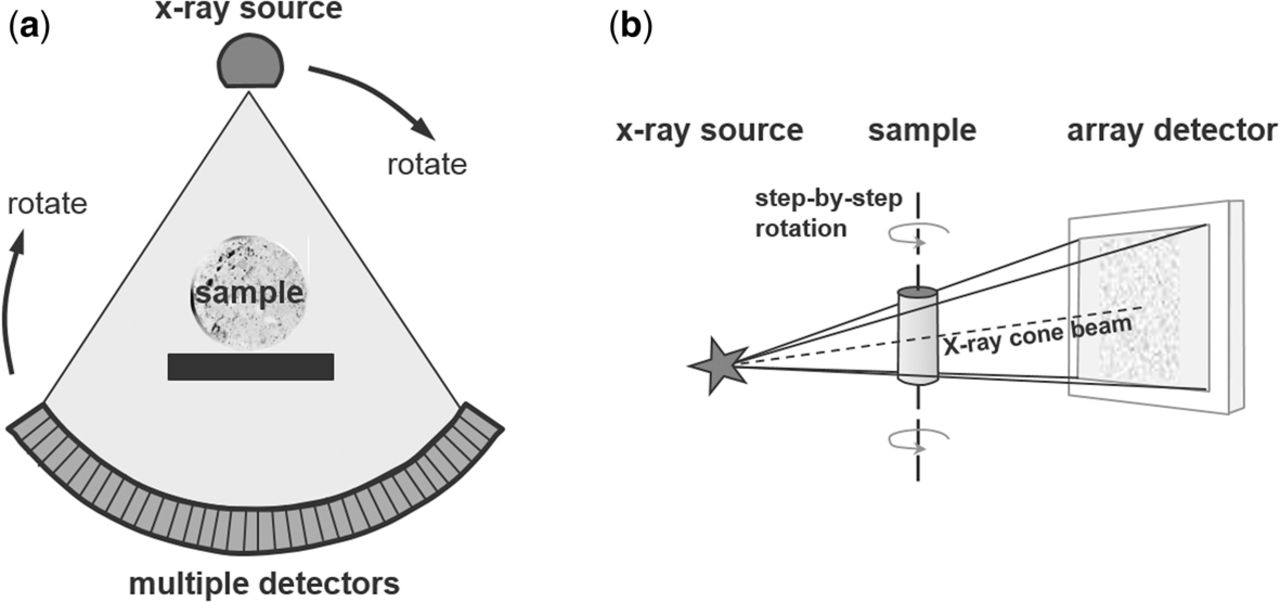}
    \caption{Computed Tomography Imaging Principle\cite{doi:10.1144/SP459.10}}
\end{figure}

A CT (Computed Tomography) scan is a medical imaging technique that uses X-rays and advanced computer algorithms to produce detailed, cross-sectional images of the body. The patient lies down on a table that slides into a donut-shaped machine, called a CT scanner. The X-ray tube, located inside the scanner, rotates around the patient, emitting a series of X-ray beams at different angles \cite{doi:10.1144/SP459.10}. As the X-ray beams pass through the body, they are absorbed by different tissues in varying degrees, creating a pattern of attenuated X-rays. The pattern of attenuated X-rays is detected by an array of detectors located opposite the X-ray tube, which converts them into electrical signals. These electrical signals are then sent to a computer that uses sophisticated algorithms to reconstruct the data into a 3D image.
\subsection{Typical Clinical Applications}
Computed axial tomography (CAT) scanning, is a medical imaging technique that has a wide range of clinical applications in medicine. Some typical clinical applications of CT imaging:
\\\\
CT scans can be used for a variety of purposes, including imaging the brain and head for abnormalities such as tumors, bleeding, or trauma. It can also be used to evaluate the sinuses, ears, and eyes. In the chest, CT scans can be used to evaluate the lungs for conditions such as pneumonia, lung cancer, and emphysema. Additionally, it can be used to evaluate the heart and blood vessels. Another specialized CT technique, called CT angiography (CTA), uses contrast material and imaging techniques to visualize blood vessels. This technique can be used to evaluate blood vessels in the head, neck, chest, abdomen, and pelvis and can also be used to evaluate for blood clots, aneurysms, and blockages in the blood vessels.
\\\\
Some advantages of CT imaging:
\\\\
CT imaging is a widely used diagnostic tool that offers several advantages. One of the major advantages of CT imaging is its high resolution, which allows it to reveal fine details of internal structures such as bones, organs, and blood vessels. This makes it useful for identifying a wide range of medical conditions. CT imaging equipment is also widely available in hospitals, clinics, and doctors' offices, making it a convenient and accessible diagnostic tool. Another advantage of CT imaging is advantageous since it does not need any kind of incisions or injections, making it relatively safe and comfortable for the patient. Furthermore, CT imaging is fast and efficient, with results usually available within a short period of time. This allows for quick identification of any issues and prompt treatment.
\\\\
However, there are also some limitations of CT imaging to consider:
\\\\
While CT imaging has many advantages, it also has some limitations and risks. One major risk associated with CT imaging is exposure to ionizing radiation, which can cause cancer and other health problems, especially in the case of multiple scans or in pregnant women. Another limitation of CT imaging is that it is limited in dynamic imaging, it can't capture dynamic processes such as blood flow or organ function, and other modalities such as fluoroscopy or angiography can be better suited for those situations. Additionally, CT scanners are relatively expensive and require a significant investment, which can be a limitation for some facilities. This may limit access to this imaging modality for some patients. It's important to balance CT scans' benefits and risks and consider alternative imaging modalities when appropriate.
\vspace{5mm}
\section{MRI and Magnetic Resonance Microscopy (MRM)}
\subsection{Basic principles}
Magnetic Resonance Imaging (MRI) and Magnetic Resonance Microscopy (MRM) are medical imaging techniques that use a magnetic field and radio waves to produce detailed images of internal structures \cite{epstein2007introduction}. The basic principles of MRI and MRM are:
\\
\\
Magnetic Resonance Imaging (MRI) and Magnetic Resonance Angiography (MRA) are advanced imaging modalities that use a powerful magnetic field to align the nuclei of hydrogen atoms in the body \cite{biswas2021ann,al2022brain}. This creates a small magnetic moment, which can be used to create detailed images of the body's internal structures. The process begins by using radiofrequency (RF) pulses to change the alignment of the hydrogen nuclei, which causes them to emit a weak radio signal. These signals are picked up by a detector and processed by a computer to create detailed images of the internal structures of the body.
\begin{figure}[h] 
    \centering

    \includegraphics[width=.6\textwidth]{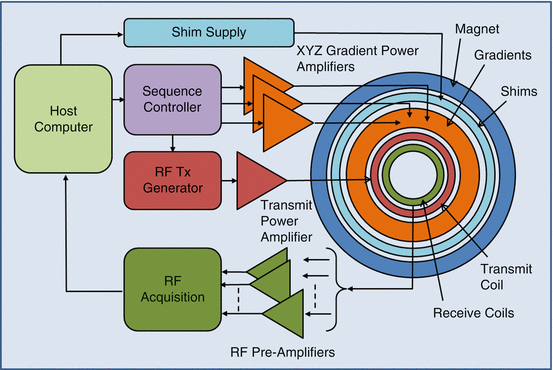}
    \caption{MRI Imaging Principle\cite{Glover2013}}
\end{figure}

MRI is a non-invasive and safe imaging technique, and it can provide detailed images of soft tissues, such as organs, muscles, and nerves\cite{Glover2013}. It is commonly used to diagnose and monitor a wide range of medical conditions, including brain and spinal cord disorders, tumors, joint and bone problems, and cardiovascular disease. When a patient enters the MRI machine, they are exposed to a strong magnetic field that aligns the hydrogen atoms in their body. Radio waves are then sent through the body, causing the aligned hydrogen atoms to produce a weak radio signal \cite{Glover2013}. The radio signal is detected by an antenna, and the data is sent to a computer for analysis. The computer uses complex algorithms to process the data and construct a detailed image of the body's internal structures. Different tissues in the body have varying levels of hydrogen atoms, which produce different signals, allowing for the differentiation between different types of tissue. MRI can produce images in different planes, such as axial, sagittal, and coronal, providing a three-dimensional view of the body.

\subsection{Typical Clinical Applications}
Magnetic Resonance Imaging (MRI) and Magnetic Resonance Microscopy (MRM) are medical imaging techniques that have a wide range of clinical applications in medicine. Some typical applications:
\\
\\
MRI and MRM are used in a variety of imaging applications to evaluate different parts of the body for abnormalities or injuries. These imaging techniques are particularly useful for evaluating the brain and spinal cord, as they can detect tumors, stroke, injury, and degenerative diseases \cite{Glover2013}. Additionally, MRI and MRM are commonly used to evaluate bones, joints, ligaments, and muscles for injuries, arthritis, or tumors. They are also used to evaluate the heart, blood vessels, and blood flow, as well as the liver, spleen, pancreas, kidneys, and other abdominal and pelvic organs. In addition to these applications, MRI is also used as a supplementary tool for mammography for breast cancer detection, especially for dense breast tissue.
\\\\
Magnetic Resonance Imaging (MRI) and Magnetic Resonance Microscopy (MRM) are medical imaging techniques that have several advantages and limitations.
\\
\\
MRI and MRM have several advantages as imaging modalities. One of the main advantages is that they do not use ionizing radiation, making them safer than other imaging modalities such as X-ray or CT. They also have the ability to provide high-resolution images of soft tissues such as muscles, tendons, and ligaments. Additionally, they can create images in multiple planes, which allows for a more complete picture of internal structures. Furthermore, MRM can provide microscopic resolution with the advantage of being non-destructive and non-invasive.\cite{TYSZKA200593} However, MRI and MRM also have several limitations. One of the main limitations is the cost, as scanners are relatively expensive and require a significant investment, which can be a limitation for some facilities. Additionally, some patients can experience discomfort or claustrophobia during the examination, which can limit the use of MRI in certain patients. Furthermore, the examination can take longer than other imaging modalities such as X-rays or CTs.
\vspace{5mm}
\section{Nuclear Imaging}
\subsection{Basic principles}
Nuclear imaging is a kind of medical imaging that creates high-resolution pictures of the body's inner workings by using minute quantities of radioactive material, called radiotracers.\cite{smith2010introduction} The basic principles of nuclear imaging are:
\\
\begin{figure}
    \centering

    \includegraphics[width=.8\textwidth]{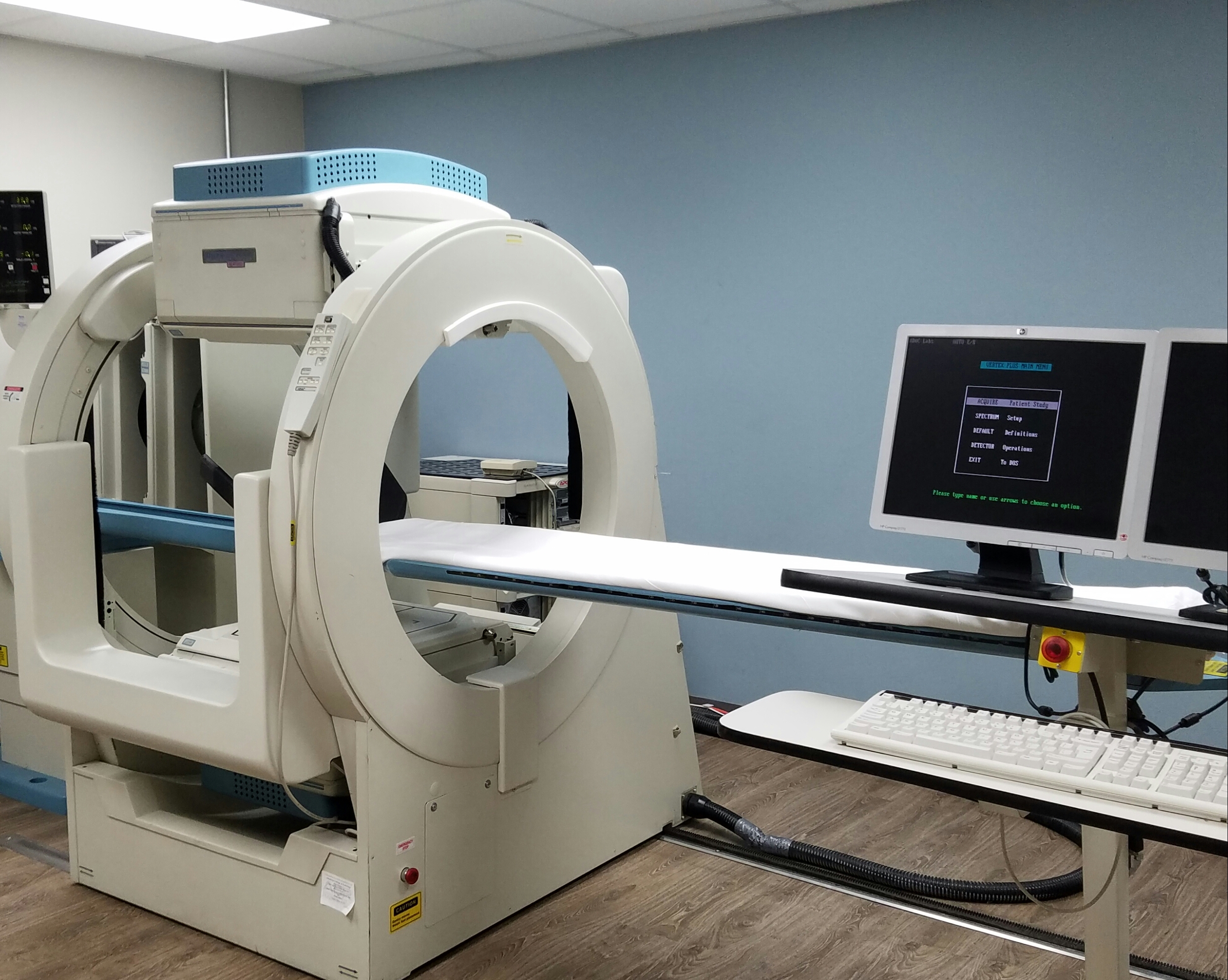}
    \caption{Ultrasound Imaging Principle\cite{smith2010introduction} }
\end{figure}

Nuclear imaging is a medical imaging technique that uses radiotracers that are introduced into the body, either by injection, inhalation, or ingestion, depending on the type of exam. The radiotracers emit gamma rays, which are detected by a special camera, called a gamma camera. This camera creates an image of the distribution of the radiotracer in the body. The gamma camera detects the gamma rays and converts them into an image, which is then processed by a computer and can be viewed in different ways, such as 2D images, 3D images, or functional images \cite{NuclearM94:online,shah2019brain}. The images obtained from nuclear imaging provide functional information about the body, such as blood flow, metabolism, or chemical activity. This information can be used to evaluate certain disorders such as cancer, inflammation, or heart function.

\subsection{Typical Clinical Applications}
Nuclear imaging is a medical imaging technique that has a wide range of clinical applications in medicine. Some typical applications:
\\
\\
Nuclear imaging is used in a variety of applications to evaluate different parts of the body. One of the main uses of nuclear imaging is cardiac imaging, where it can be used to evaluate blood flow to the heart and detect any blockages or abnormalities, such as coronary artery disease or heart failure. Additionally, it is used in cancer imaging, especially in cases where traditional imaging modalities like X-ray or CT is not as effective to detect and staging cancer. Nuclear imaging can also be used to detect bone disorders such as osteoporosis or bone cancer and to evaluate bone healing after injury or surgery. Furthermore, it can be used to evaluate the function of the thyroid gland and detect any abnormalities such as thyroid cancer or hyperthyroidism.
\\\\
Nuclear imaging is a medical imaging technique that has several advantages and limitations.
\\
\\
Nuclear imaging has several advantages as a medical imaging modality. One of the main advantages is its high sensitivity, which allows it to detect small tumors or abnormalities that can not be visible on other imaging modalities. Additionally, it can create images in multiple planes, which allows for a more complete picture of internal structures. Nuclear imaging is also non-invasive, requiring no incisions or injections, making it relatively safe and comfortable for the patient. Furthermore, it provides functional information about the body, such as blood flow, metabolism, or chemical activity, which can be used to evaluate certain disorders such as cancer, inflammation, or heart function. Positron Emission Tomography (PET) is a type of nuclear imaging that uses short-lived radionuclides to create detailed functional images of the body \cite{Positron67:online}, it's used to evaluate a wide range of conditions such as cancer, heart disease, brain disorders, and others.
\\
\\
However, nuclear imaging also has several limitations. One of the main limitations is the cost, as nuclear imaging equipment is relatively expensive and requires a significant investment, which can be a limitation for some facilities \cite{nasim2021prominence}. Additionally, nuclear imaging requires specialized equipment and trained personnel, which can limit its availability in some areas. Furthermore, nuclear imaging involves exposure to ionizing radiation, which can cause cancer and other health problems \cite{NuclearM67:online}, especially in the case of multiple scans or in pregnant women.
\vspace{5mm}
\section{Ultrasound Imaging}
\subsection{Basic principles}
Ultrasound imaging is a medical imaging technique that uses high-frequency sound waves to produce detailed images of the internal structures of the body. The basic principles of ultrasound imaging are:
\\
\\
Ultrasound imaging is a medical imaging technique that uses a transducer, a device that emits high-frequency sound waves and detects the echoes of these waves as they bounce back from internal structures. The sound waves are sent into the body and as they hit a boundary between different types of tissue, some of the sound waves are reflected back to the transducer. These echoes are picked up by the transducer and converted into electrical signals, which are then processed by a computer to create detailed images of the internal structures of the body. These images can be used to evaluate organs, blood vessels, and fetuses during pregnancy, among other things.
\begin{figure}
    \centering

    \includegraphics[width=.8\textwidth]{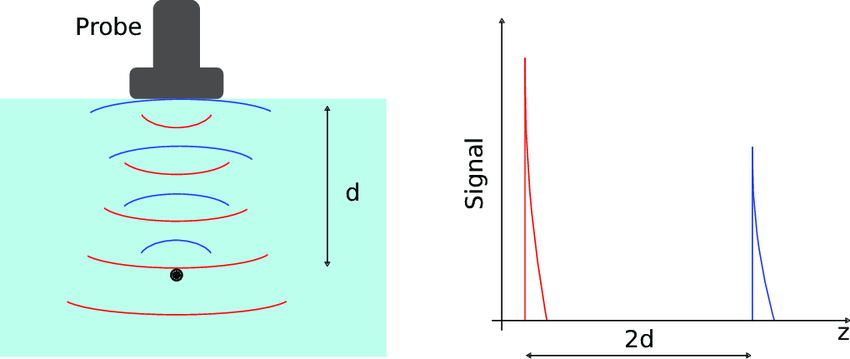}
    \caption{Ultrasound Imaging Principle\cite{phdthesis}}
\end{figure}

A diagram of the ultrasound imaging method's basic principles can be found in Figure 5 \cite{phdthesis}. It gives depth information that is then used to create a two- or three-dimensional image of the sample based on the measurement of the time delay between the red-coloured ultrasound wave that is being emitted and the blue-coloured waves that are being reflected back from each layer of the sample.

\subsection{Typical Clinical Applications}
Ultrasound imaging is a widely used medical imaging technique that has a range of clinical applications. Some typical applications include:
\\
\\
Ultrasound imaging is used for a variety of applications to evaluate different parts of the body. One of the main uses of ultrasound imaging is abdominal imaging, where it is used to evaluate the liver, gallbladder, pancreas, spleen, and kidneys. Additionally, it is commonly used in musculoskeletal imaging, to evaluate muscles, tendons, ligaments, and joints to detect injury, inflammation, or tumors \cite{Xie:2021:2156-7018:930}. Ultrasound imaging is also used to evaluate the thyroid gland and detect any abnormalities such as nodules or cysts. Furthermore, it's commonly used in Obstetrics and Gynecology, to evaluate pregnancy, including the detection of the fetus, placenta, and amniotic fluid, as well as to evaluate the female reproductive organs, such as the ovaries, uterus, and fallopian tubes.
\\\\
Ultrasound imaging is a medical imaging technique that has several advantages and limitations.
\\
\\
Ultrasound imaging has several advantages as a medical imaging modality. One of the main advantages is that the equipment is portable, which allows for imaging to be performed at the patient's bedside or in remote locations. Additionally, ultrasound equipment is relatively inexpensive, which makes it widely available in medical facilities. It is also preferable to other imaging modalities like X-ray and CT since it does not use harmful ionizing radiation. Furthermore, it can create images in real-time, which allows for the dynamic evaluation of internal structures and functions.
Image segmentation has been addressed in \cite{karim2023c} which has numerous applications in medical imaging
\\
\\
However, ultrasound imaging also has several limitations. One of the main limitations is the limited penetration depth, which can make it difficult to image deep structures or structures behind bone or gas. Furthermore, aberrations such as bone or air echoes can distort pictures, making it hard to discern interior structures. Additionally, the ultrasound operator's degree of knowledge might alter the quality of the ultrasound pictures and the accuracy of the diagnosis.
\vspace{5mm}

\section{Emerging Technologies for in Vivo Imaging}

\subsection{Electrical Impedance Tomography (EIT)}
\subsubsection{Basic principles}
Electrical Impedance Tomography (EIT) is a medical imaging technique that uses electrical currents to produce images of the interior of the body. The basic principles of EIT are:
\\
\\
EIT (Electrical Impedance Tomography) is a medical imaging technique that uses a small number of electrodes placed on the surface of the body. These electrodes are used to apply small electrical currents to the body and to measure the resulting voltage changes. When the electrical currents pass through the body, they encounter different types of tissue with different electrical properties, such as conductivity and permittivity. These properties affect the distribution of the electrical current, which can be used to produce images of the internal structure of the body. Electrodes measure electrical activity in various parts of the body, and the computer uses this information to build representations of organs including the lungs, heart, and brain \cite{Costa2009}. EIT is a non-invasive, radiation-free imaging technique that has the potential to be used in various medical applications, such as lung imaging, brain imaging, and breast imaging.
\begin{figure}
    \centering

    \includegraphics[width=.8\textwidth]{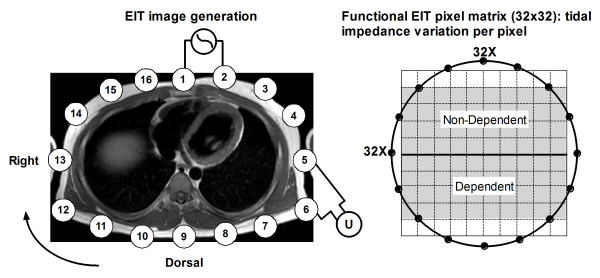}
    \caption{EIT Imaging Principle\cite{article}}
\end{figure}

The functional EIT image (fEIT) and the electrical impedance tomography (EIT) principle (Figure 6). Between pairs of nearby surface electrodes (1 to 16), electrical excitation currents are applied; the resulting voltages are monitored between the other electrodes (U) \cite{article}. The impedance change brought on by the tidal volume is separated into a 32 by 32 matrix in the fEIT image. An image of the ventilation distribution is produced because each pixel carries a unique tidal impedance variation. In the right panel, the ROIs that are ventral to dorsal are highlighted in gray.
\subsubsection{Limited Typical Clinical Usage}
Electrical Impedance Tomography (EIT) is a relatively new medical imaging technique that is still being researched and developed, and its clinical applications are still being explored. However, some potential clinical applications include:
\\\\
EIT is a promising imaging modality that has been explored in various medical applications. One of the main applications of EIT is in lung imaging, where it has been shown to have the potential to evaluate lung function and detect changes in lung structure, for instance, among those who suffer from respiratory disorders like COPD or ARDS (Acute Respiratory Distress Syndrome). Additionally, EIT has been proposed as a method to evaluate heart function, such as cardiac output and blood flow, which could be useful for patients with heart failure or cardiac arrhythmias. Furthermore, it has been proposed as a method to evaluate brain function, such as cerebral blood flow, which could be useful for patients with stroke, brain injury, or brain tumors. Lastly, EIT has been proposed as a method to evaluate the function of the gastrointestinal tract, such as motility, which could be useful for patients with gastrointestinal disorders.
\\
\\
However, it's important to note that EIT is still a new technology and there is still ongoing research to explore its potential further.
\\\\
Electrical Impedance Tomography (EIT) is a medical imaging technique that has several advantages and limitations.
\\
\\
EIT is a medical imaging modality that has several advantages. One of the main advantages is that the equipment is portable, which allows for imaging to be performed at the patient's bedside or in remote locations. Additionally, EIT equipment is relatively inexpensive, which makes it widely available in medical facilities. It is also preferable to other imaging modalities like X-ray and CT since it does not use harmful ionizing radiation. In addition, it does not need any incisions or injections, thus it is a generally safe and pleasant option for the patient. On the other hand, EIT has several serious drawbacks. The inability to see through bone or through gas might be problematic when trying to scan deeper structures. Additionally, images can be affected by interference from other sources of electrical activity, such as muscle activity or electronic devices. Furthermore, the quality of EIT images can vary depending on the skill of the operator, and the level of expertise can affect the accuracy of the images and the diagnosis.

\subsection{Advancements and new modalities}
Several in vivo imaging methods are in development. These technologies might enhance medical imaging and provide new body assessment methods. New technologies include:
\\
\\
There are several new imaging modalities that are being developed to produce high-resolution images of the internal structures of the body\cite{Weissleder2001}. Photoacoustic imaging creates high-resolution pictures of bodily structures using laser light and ultrasound. It has potential applications in cancer detection, vascular imaging, and optoacoustic imaging. Another technology is Optical Coherence Tomography (OCT), which uses infrared light to produce high-resolution images. It has potential applications in ophthalmology, cardiology, and oncology. Magnetomotive Optical Coherence Tomography (MM-OCT) combines magnetic fields with optical coherence tomography to provide pictures of anatomical components within the body. Possible uses include vascular imaging and the detection of malignancy. Another method that combines lasers and ultrasound to provide quantitative pictures of the body's interior architecture is called quantitative photoacoustic tomography (QPAT). It might be used for things like functional imaging and cancer diagnosis. Another method that employs infrared light to create pictures of the body's interior architecture is diffuse optical tomography (DOT). Both cancer and functional brain imaging might benefit from this discovery. 
\\
\\
Each of these technologies has its own advantages and disadvantages and the selection of the appropriate technique depends on the clinical situation and the patient's condition.
\vspace{5mm}
\newpage
\section{Comparative Analysis}
\begin{table}[]
\caption{Comparative Analysis}
\resizebox{\textwidth}{!}{%
\begin{tabular}{ | p{2cm} | p{3cm} | p{4cm} | p{2cm} | p{2cm} |}
\hline
Modality                        & Working Principle                                                                              & Applications                                                                                                                              & Advantages                                            & Limitations                                                                     \\ \hline
X-ray                           & Using ionizing radiation to produce images of the internal structure of a body                 & Detecting broken bones, monitoring treatment of conditions such as pneumonia, monitoring the healing of fractures                      & Inexpensive, widely available, quick results          & Low-resolution images, ionizing radiation exposure                              \\ \hline
CT Scan                         & X-ray technology combined with computer processing to produce detailed images                  & Detecting cancers, identifying blood clots, assessing organ damage, diagnosing spinal problems                                            & High-resolution images, non-invasive                  & Ionizing radiation exposure, high cost                                          \\ \hline
MRI                             & Combination of powerful magnetic fields and radio waves allows for the creation of high-resolution photographs of hidden structures & Detecting tumors, brain and spinal cord injuries, joint problems, and monitoring the progression of conditions such as multiple sclerosis & Non-ionizing radiation, detailed images               & Long examination time, high cost, not suitable for patients with metal implants \\ \hline
Ultrasound Imaging              & Using high-frequency sound waves to produce images                                             & Monitoring the growth and development of a fetus, evaluating organs and tissues, detecting tumors and cysts                               & Non-invasive, no ionizing radiation exposure          & Operator dependent, limited view of deep structures                             \\ \hline
Nuclear Imaging                 & Using radioactive tracers to produce images                                                    & Detecting diseases and conditions such as cancer, heart disease, and neurological conditions                                              & High specificity for certain conditions, non-invasive & Limited view of the structure, exposure to ionizing radiation                       \\ \hline
Electrical Impedance Tomography & Using electrical currents to produce images                                                    & Monitoring changes in tissue, measuring organ function                                                                                    & Non-invasive, portable                                & Limited spatial resolution, operator dependent                                  \\ \hline
Cardiovascular Imaging          & Various techniques to produce images of the heart and blood vessels                            & Diagnosing heart disease, monitoring treatment                                                                                            & High-resolution images, non-invasive                  & High cost, operator dependent                                                   \\ \hline
\end{tabular}%
}

\end{table}

The functioning principles, uses, benefits, and limitations of each modality are compared in this table (Table 1) along with their advantages and disadvantages.

\section{Specialized Techniques}
\subsection{Contrast-Enhanced MRI}
Contrast-enhanced magnetic resonance imaging (MRI) employs a contrast agent to highlight bodily structures. Before the MRI, a gadolinium-based contrast agent is injected into the patient's bloodstream. MRI scanners can detect magnetic field changes caused by the contrast agent once it reaches the region of interest \cite{Lohrke2016}. This helps see blood arteries, tumors, and inflammation.
\\\\
The contrast agent reduces target tissue water proton T1 relaxation time. This enhances T1-weighted imaging and distinguishes healthy from diseased tissue.
\\\\
Contrast-enhanced MRI is used to evaluate a wide range of medical conditions \cite{https://doi.org/10.1002/nbm.2940}, including:

\begin{itemize}
    \item Cancer diagnosis and staging
    \item Evaluating blood vessels and blood flow
    \item Evaluating the liver, kidney, and other organs
    \item Evaluating the brain and spinal cord
\end{itemize}

There are dangers connected with using the contrast agent, such as an allergic response or renal damage, thus not all individuals are good candidates for contrast-enhanced MRI. Before deciding to schedule a checkup, your physician will assess the benefits and risks.
\subsection{MR Approaches for Osteoarthritis}
Magnetic resonance imaging (MRI) is a powerful tool for evaluating osteoarthritis, a condition characterized by the degeneration of the cartilage in joints. There are several MR approaches that can be used to evaluate osteoarthritis:
\\\\
Magnetic Resonance Imaging (MRI) is a powerful imaging modality that can be used to evaluate the cartilage and detect early signs of osteoarthritis \cite{doi:10.1148/radiol.2262012190}. T1-weighted imaging is one technique that is sensitive to changes in the water content of the cartilage and can be used to detect early signs of osteoarthritis, such as cartilage loss, changes in the bone marrow, or the presence of bone spurs. Another technique is T2-weighted imaging, which is sensitive to changes in the biochemical composition of the cartilage and can be used to detect later stages of osteoarthritis, such as increased water content, increased proteoglycan loss, or increased collagen content. Sodium MRI is a new technique that uses sodium ions to detect changes in the cartilage, it can detect changes in the matrix of the cartilage and bone marrow edema, which is indicative of early osteoarthritis \cite{HAYASHI2016161}. Another technique is Dynamic contrast-enhanced MRI, which can be used to evaluate the blood flow in the joint, which can be useful in detecting inflammation or synovitis. Each technique has its own advantages and disadvantages and the selection of the appropriate technique depends on the clinical situation and the patient's condition.
\\\\
It's important to keep in mind that MRI is a diagnostic tool and the results should be interpreted by a radiologist in conjunction with clinical examination and other tests.
\subsection{Cardiovascular Imaging}
Cardiovascular imaging is a subspecialty of medical imaging that focuses on the evaluation of the heart and blood vessels \cite{doi:10.1016/j.jcmg.2009.11.010}. Several imaging modalities are used in cardiovascular imaging, including:
\\\\
There are several imaging modalities that can be used to evaluate the blood vessels and the heart. Computed Tomography Angiography (CTA) uses X-rays and computer processing to produce detailed images of the blood vessels and is used to detect blockages or aneurysms in the coronary, carotid, and peripheral vessels. Another modality is Magnetic Resonance Angiography (MRA) which uses magnetic fields and radio waves to produce detailed images of the blood vessels and is also used to detect blockages or aneurysms in the coronary, carotid, and peripheral vessels \cite{MachineL5:online, doi:10.1016/j.jacc.2018.12.054}. 
Both CTA (Computed Tomography Angiography) and MRA (Magnetic Resonance Angiography) are imaging techniques used to visualize blood vessels in the body. However, CTA uses X-rays to create detailed images of blood vessels, while MRA uses magnetic fields and radio waves to produce images. Additionally, CTA may expose patients to ionizing radiation, while MRA does not use ionizing radiation.
Another modality is Cardiac Magnetic Resonance Imaging (CMR) which uses magnetic fields and radio waves to produce detailed images of the heart and blood vessels and is used to evaluate the heart's structure and function and detect heart diseases such as cardiomyopathies or valvular diseases. Echocardiography is another modality that uses ultrasound waves to produce detailed images of the heart and blood vessels, it is also used to evaluate the heart's structure and function and detect heart diseases such as cardiomyopathies or valvular diseases. Each modality has its own advantages and disadvantages and the selection of the appropriate modality depends on the clinical situation and the patient's condition.
\\\\
The use of ML and AI has taken these modalities to new heights. These imaging modalities can be used alone or in combination to provide a comprehensive evaluation of the cardiovascular system, and to detect and diagnose cardiovascular diseases.
\subsection{Medical Imaging Data Mining and Search}
Medical imaging data mining and search are techniques that are used to extract useful information from large amounts of medical imaging data \cite{Wang2016/05}. This information can be used to improve patient care, to support research, and to improve the efficiency of the healthcare system. Also has its own limitations as well\ cite{884767}. Following are some techniques \cite{10.1007/978-3-642-31965-5_15}:

\begin{enumerate}
    \item Content-based Image Retrieval (CBIR): CBIR is a technique that uses image features such as color, texture, or shapes to search for similar images in a large database of medical images \cite{shukran2021new}. This technique can be used to find images of similar anatomy, pathology, or disease progression.
    
    \item Text-based Search: Text-based search allows to search images based on associated metadata like patient's demographics, diagnosis, anatomic location, etc \cite{Aggarwal_2020_WACV}.
    
    \item Machine learning: Machine learning is a technique that uses algorithms to automatically extract features from images and classify them into different categories. This technique can be used to identify specific features of a disease, such as a tumor, or to predict patient outcomes \cite{kaur2019systematic}.
    
    \item Natural Language Processing (NLP): NLP is a technique that uses algorithms to extract information from unstructured text data, such as radiology reports, to identify relevant patient information and to support decision-making \cite{leeson2019natural}.
    
    \item Radiomics: Radiomics is the process of extracting quantitative information from medical images using mathematical algorithms. It enables the extraction of information from medical images that would otherwise be invisible to the human eye and can be used to predict patient outcomes or to evaluate treatment response \cite{hatt2019machine}.
    
\end{enumerate}
These techniques can be used to improve patient care by providing physicians with more accurate and detailed information about a patient's condition, to support research by providing a large amount of data for analysis, and to improve the efficiency of the healthcare system by reducing the need for repeat imaging and helping to identify patients at risk of certain diseases \cite{Wang2016/05}.
\vspace{5mm}

\section{Discussion and Conclusion}

X-ray imaging is one of the oldest and most widely used medical imaging techniques. It is commonly used for the diagnosis of bone fractures and other conditions affecting the skeleton. However, its use is limited in the visualization of soft tissue structures.
\\\\
CT scans use X-rays and computer algorithms to generate cross-sectional images of the body. This modality is widely used for the diagnosis of conditions affecting the brain, chest, abdomen, and pelvis. However, CT scans use ionizing radiation, which may increase the risk of cancer.
\\\\
MRI uses a strong magnetic field and radio waves to generate detailed images of the body's internal structures. It is widely used for the diagnosis of conditions affecting the brain and spine, as well as joints and muscles. MRI is a safe and non-invasive alternative to CT scans, but it is often more expensive and time-consuming.
\\\\
Ultrasound imaging uses high-frequency sound waves to produce images of the body's internal structures. It is widely used for the diagnosis of conditions affecting the abdomen, pelvis, and reproductive system. Ultrasound imaging is safe and non-invasive, but its resolution may be limited in certain applications.
\\\\
Nuclear medicine imaging involves the use of radioactive isotopes to produce images of the body's internal structures. It is widely used for the diagnosis of cancer and other conditions affecting the bones and organs. The amount of radiation exposure during nuclear medicine imaging is generally low and comparable to the amount of radiation exposure received during a standard X-ray. The radiopharmaceuticals used in nuclear medicine imaging are designed to be quickly eliminated from the body, minimizing the amount of radiation exposure.
\\\\
Cardiovascular imaging is a specialized field that uses various imaging techniques to visualize the heart and blood vessels. And, techniques such as echocardiography, angiography, and magnetic resonance angiography. Cardiovascular imaging is essential for the diagnosis and treatment of cardiovascular diseases.

The results of this review demonstrate the importance of medical imaging modalities in the diagnosis and treatment of various diseases. Each modality has its strengths and limitations, and the choice of modality should be based on the specific needs of each patient. Because of the radiation risk associated with CT and X-rays, MRI is not used. They use a different imaging process to produce image contrast. Therefore, while MRI is helpful for detecting some diseases, CT is good for detecting others.  
\\\\
The results of this review also highlight the need for continued research and advancements in the field of medical imaging. The development of new and improved modalities will play a critical role in the early detection and treatment of diseases. The increasing availability of specialized imaging techniques, such as Electrical Impedance Tomography and Cardiovascular Imaging, has improved the accuracy of diagnoses and treatment plans.

In conclusion, Medical imaging modalities are a crucial tool in modern healthcare and play a vital role in the diagnosis, management, and treatment of various diseases. There are several different modalities available, each with its own advantages and limitations. X-ray, Computed Tomography (CT), Magnetic Resonance Imaging (MRI), Nuclear Imaging, Ultrasound, Electrical Impedance Tomography (EIT), and Emerging Technologies for in vivo Imaging are the common medical imaging modalities.
\\\\
In addition to these modalities, there are also several advanced techniques, such as contrast-enhanced MRI, MR approaches for osteoarthritis, Cardiovascular Imaging, and Medical Imaging Data Mining and Search, that can provide additional information and improve the accuracy of diagnosis.
\\\\ 
However, it's important to keep in mind that medical imaging is not without risks and limitations and the results should be interpreted by a radiologist or physician in conjunction with clinical examination and other tests.

\bibliography{main.bib}
\bibliographystyle{unsrt}

\end{document}